\documentclass[12pt]{article}
\usepackage{amsfonts}
\usepackage{amsmath}
\usepackage{mathrsfs}
\usepackage{color}
\usepackage{tikz}
\usepackage{authblk}
\usepackage{mathrsfs,amscd,amssymb,amsthm,amsmath,bm,graphicx,psfrag,subfigure,url}

\usepackage[linesnumbered,ruled,vlined]{algorithm2e}
\SetKwInOut{Input}{input}
\SetKwInOut{Output}{output}

\usetikzlibrary{positioning, shapes.geometric}

\allowdisplaybreaks
\usepackage{indentfirst}

\def \[{\begin{equation}}
\def \]{\end{equation}}

\newtheorem{theorem}{Theorem}[section]

\newtheorem{definition}{Definition}

\newtheorem{case}{Case}

\newtheorem{lemma}[theorem]{Lemma}

\begin{document}
\title{\bf A constant time complexity algorithm for the unbounded knapsack problem with bounded coefficients}
\author{Yang Yang\thanks{Email: zhugemutian@outlook.com}}
\affil{School of Mathematical Sciences, Xiamen
University, Xiamen 361005, P.R. China}
\date{}
\maketitle

\begin{abstract}
Benchmark instances for the unbounded knapsack problem are typically generated according to specific criteria within a given constant range $R$, and these instances can be referred to as the unbounded knapsack problem with bounded coefficients (UKPB). In order to increase the difficulty of solving these instances, the knapsack capacity $C$ is usually set to a very large value. Therefore, an exact algorithm that neither time complexity nor space complexity includes the capacity coefficient $C$ is highly anticipated.

In this paper, we propose an exact algorithm with time complexity of $O(R^4)$ and space complexity of $O(R^3)$. The algorithm initially divides the multiset $N$ into two multisubsets, $N_1$ and $N_2$, based on the profit density of their types. For the multisubset $N_2$ composed of types with profit density lower than the maximum profit density type, we utilize a recent branch and bound (B\&B) result by Dey et al. (Math. Prog., pp 569-587, 2023) to determine the maximum selection number for types in $N_2$. We then employ the Unbounded-DP algorithm to exactly solve for the types in $N_2$. For the multisubset $N_1$ composed of the maximum profit density type and its counterparts with the same profit density, we transform it into a linear Diophantine equation and leverage relevant conclusions from the Frobenius problem to solve it efficiently. In particular, the proof techniques required by the algorithm are primarily covered in the first-year mathematics curriculum, which is convenient for subsequent researchers to grasp.
\end{abstract}

\vspace{2mm} \noindent{\bf Keywords}: unbounded knapsack problem with bounded coefficients; dynamic programming; branch and bound; subset sum problem; linear Diophantine equation; Frobenius problem
\vspace{2mm}

\setcounter{section}{0}

\section{Introduction}

\begin{definition}
\cite{MT1990,KPP2004,CILM2022a,CILM2022b,BB2019}
Given a multiset $N$ consisting of $n'$ item types and a total of $n$ items, where all items of the $j$-th type have same profit of $p_j$ and weight of $w_j$, and $p_j,w_j\in \mathbb{N}^+$. Each type can be selected unlimited times. The goal of the unbounded knapsack problem (UKP) is to find an optimal multisubset(i.e., a set that allows multiple copies of the same element) of the multiset $N$ such that its total profit is maximized without exceeding the knapsack capacity $C$. The problem can be formulated as follows.
\begin{align}
\max\ & f({\bf X})=\sum\limits_{j=1}^{n'} p_jx_j\label{UKP-model-obj}\\
\text{s.t.}\quad & \sum\limits_{j=1}^{n'}w_jx_j\le C,\label{UKP-model-con-1}\\
& x_j\in\mathbb{N}. \label{UKP-model-con-2}
\end{align}
The variable $x_j$ denotes the frequency with which the $j$-th type is selected. The UKP is akin to the 0-1 Knapsack Problem (0-1 KP), sharing the same objective and capacity constraints. A key distinction of the UKP is that the $j$-th type can be selected an unlimited times, as detailed in Equation (\ref{UKP-model-con-2}). Specifically, without loss of generality, we arrange items in descending order of their profit density ratio, i.e., $p_1/w_1 \geq p_2/w_2 \geq \dots \geq p_{n'}/w_{n'}$.
\end{definition}

The UKP is not only a variant of the 0-1 KP, but it is also a special form of integer linear programming when the constraint number is only 1 (i.e., $m=1$). As such, it is an $\mathcal{NP}$-hard problem
\cite{Papadimitriou1981,SL2010,Wolsey2020,CCZ2014}. $\mathcal{NP}$-hard problems have widespread applications in commercial decision-making and industrial manufacturing, leading to significant economic value. Therefore, efficient and exact algorithms have attracted the attention of numerous researchers. Unfortunately, the research progress on efficient algorithms is slow due to limitations in mathematical tools\cite{SBR2002,DS2005}. Currently, there is no exact polynomial-time algorithm to solve the UKP, except for dynamic programming (DP), which can provide an exact solution in pseudo-polynomial time complexity\cite{GJ1979,Bellman1957}.

Although there are currently no efficient algorithms to solve general $\mathcal{NP}$-hard problems exactly, taking into consideration practical application context, extensive research has been conducted on improving algorithms in the following three directions. The first direction is based on methods such as linear relaxation within the framework of the branch and bound algorithm (B\&B)\cite{AA1960}. By improving the Dantzig bound\cite{Dantzig1957} and later the cutting plane method\cite{CCZ2014,GJ2003,LSW2015}, it aims to solve as many decision variables as possible in polynomial time complexity, reducing the number of exhaustive subproblems and enhancing the algorithm's solving speed. The selection of upper bounds is typically within a time complexity of $O(n^2)$, such as in\cite{Pisinger1997,MPT1999,BI2019,GMMT2020}. The second direction utilizes mathematical tools, including number theory, to enhance the speed of the basic dynamic programming(BDP) and reduce the pseudo-polynomial time complexity, such as in
\cite{MT1984,Bringmann2017,KX2019,DMZ2023}. The third direction involves adding as few and as general constraints as possible to transform $\mathcal{NP}$-hard problems into $\mathcal{P}$, such as utilizing specific data features\cite{MNT1975,HL1976,ZJNW2001,DW2011,ABS2016} or fixing the number of variables\cite{HW1976,Kannan1980,Scarf1981,Lenstra1983,Chenetal2008}.

To demonstrate the performance of improved algorithms, researchers typically test them on benchmark instances of varying scales, which are generated according to specific criteria. Given a positive integer $R$, it raises an intriguing question of whether an instance with parameters ranging from 1 to $R$ can be solved in polynomial time. Taking the UKP as an example, if the profit $p_j$ and weight $w_j$ of each type in an instance are positive integers that do not exceed $R$, the number of possible decisions is limited to at most $R^2$. To incorporate bounded coefficients, the Unbounded Knapsack Problem with Bounded coefficients (UKPB) introduces the following constraints to the UKP model:
\begin{align}
p_j,w_j\le R. \label{UKP-model-con-3}
\end{align}

In particular, dynamic programming can solve the 0-1 KP exactly within polynomial time complexity $O(n^2R)$ when inequality (\ref{UKP-model-con-3}) holds for any capacity constraint $C$. whereas, for any capacity constraint $C$ in the UKPB, the input length can be expressed as $\log_2 C$. The BDP solves the UKPB exactly in time complexity $O(nC)$, while basic B\&B solves it in time complexity $O(\prod_{j=1}^{n'}\frac{C}{w_j})$. Clearly, when $C$ is a sufficiently large number, neither BDP nor basic B\&B can solve the UKPB exactly within polynomial time complexity. It is important to note that the polynomial-solvable algorithm is not straightforward in its design and implementation. However, since the fixing number of variables, a polynomially solvable algorithm for integer linear programming, and therefore also for the UKPB, was presented in the famous paper by Lenstra \cite{Lenstra1983,KPP2004}, with the proof technique derived from the geometry of numbers. Sadly, due to the hardness of the proof techniques, it is worth expecting simpler and more efficient algorithms.
\subsection{Our results}
In this paper, we propose an exact and simple pseudo-polynomial time complexity algorithm for the UKPB. The time and space complexities of the algorithm are $O(R^4)$ and $O(R^3)$ respectively, where $R$ is an upper bound of coefficients. These complexities do not change with the capacity of the knapsack $C$. In particular, since $R$ is a given integer, the time and space complexities of the algorithm are both $O(1)$. Therefore, the algorithm is a constant time complexity algorithm.

Furthermore, the method primarily employs fundamental mathematical concepts that are typically included in the first-year mathematics curriculum. The proof tools employed are straightforward, ensuring subsequent researchers can easily grasp and build upon.

\subsection{Sketch of proof techniques}

Our method first obtains the break type $b$ and residual capacity $r$ using the common greedy algorithm \cite{KPP2004,BB2019}. Then, the types in the set $N$ are divided into two disjoint subsets, $N_1$ and $N_2$, where the profit density of types in $N_1$ and $N_2$ are equal to and less than the break type, respectively. That is, $N_1 = \{j | p_j/w_j = p_1/w_1\}$ and $N_2 = \{j | p_j/w_j < p_1/w_1\}$. Finally, through linear relaxation, we obtain an upper bound for the problem.

For the types in $N_2$, we employe a recent B\&B theoretical result by Dey et al.\cite{DDM2023}, which derived from analyzing the performance of the B\&B in solving random integer linear programming problems, and obtain an upper bound on the number of times a type in $N_2$ can be selected. In other words, for the $j$-th type in $N_2$, if a solution has $x_j > R^2$, then the solution cannot be optimal. Consequently, types in $N_2$ can be exactly solved with a space complexity of $O(R^3)$ and a time complexity of $O(R^4)$ using the Unbounded-DP method \cite{KPP2004}.

For the types in $N_1$, we first utilize the Unbounded-DP to solve the subproblem with a capacity of $2R^2+2R$ and decision variables from the types in $N_1$. When the knapsack capacity $C$ exceeds $2R^2+2R$, we transform the problem of finding the optimal types selection into a Linear Diophantine Equation(LDE) \cite{Greenberg1980,CC1982,DM1988,IR1990,CD1994,Rosen2005,Alfonsin2005}. We employ the Euclidean Algorithm to compute the greatest common divisor of the types weight in $N_1$, denoted as $\text{gcd}(W_1)$, which can be computed with a time complexity of $O(n\log R)$. Subsequently, we apply B$\acute{e}$zout's theorem and classical results of Frobenius numbers \cite{Alfonsin2005,Marklof2010,AHH2011,FS2011} to obtain an optimal solution for the problem.

\subsection{Organization of the paper}

The rest of this paper is organized as follows: Section 2 provides an overview of the preliminary and tools required for the subsequent proofs. The main theoretical results and their proofs are presented in Section 3. In the final section, we summarize the paper.

\section{Preliminary}

\subsection{Dynamic programming}

For $\mathcal{NP}$-hard problems that satisfy the Bellman Optimality Principle, dynamic programming (DP) has been an important method to solving them. BDP, which typically takes time and space complexity respectively of both $O(nC)$, yields the optimal solution to the UKP, denoted by $\textbf{Y}$ \cite{ KPP2004}. Let $\text{DP}(j,k)$ represent the optimal solution for selecting the first $j$ items under the constraint of a given knapsack capacity of $k$. Clearly, $\text{DP}(n,C)$ is the optimal solution to the original problem and can be obtained through the following recursive equation
\begin{align}\label{BDP}
\begin{split}
\text{DP}(j,k)= \left \{
\begin{array}{ll}
\text{DP}(j-1,k), & \text{if } k<w_j,\\
\max\{\text{DP}(j-1,k), \text{DP}(j-1,k-w_j)+p_j\} & \text{if } k\ge w_j.
\end{array}
\right.
\end{split}
\end{align}

BDP requires $n$ arrays of size $C$ for a given capacity constraint. As the capacity increases, the storage cost increases significantly. With the development of data structures, a Unbounded Dynamic Programming (Unbounded-DP) approach based on linked list storage and addressing unbounded constraints has been proposed in \cite{KPP2004}. Although the time complexity remains unchanged, the space complexity of the Unbounded-DP can be reduced to $O(n'+C)$ which is a significant improvement. The pseudo-code for the Unbounded-DP is presented as follows.

\begin{algorithm}[]\label{Unbounded-DP}
    \caption{Unbounded-DP}
    \Input{$n', W, P, C$}
    \Output{$\textbf{Y}$}
    $\text{DP} = zeros(1,C+1)$, $\text{T}=\text{DP}$\\
    \For{$j = 1:n'$}{
        \For{$k = w_j+1:C+1$}{
            \If{$\text{\normalfont DP}(k-w_j)+p_j> \text{\normalfont DP}(k)$}{
                $\text{DP}(k)=\text{DP}(k-w_j)+p_j$, $\text{T}(k) = j$
            }
        }
    }
    $t = C+1$, $\textbf{Y}=zeros(1,n')$\\
    \While{$t>1$}{
        $k=\text{T}(t)$, $y_k = y_k+1$, $t = t-w_k$
    }
\end{algorithm}

The Unbounded-DP algorithm first initializes a recursive array $\text{DP}$ of length $C+1$ and a types selection array $\text{T}$. Next, the algorithm employs a recursive method similar to BDP to obtain the array $\text{DP}$ (Alg. \ref{Unbounded-DP}, lines 2-5). Clearly, the objective function value is $f(\textbf{Y})=\text{DP}(C+1)$. Finally, the optimal solution $\textbf{Y}$ is obtained from the array $\text{T}$ (Alg. 1, lines 7-8).

It is worth noting that for the UKP, even state-of-the-art dynamic programming algorithms such as EDUK have a time complexity of $O(nC)$\cite{KPP2004,BB2019,APR2000}. Furthermore, the most special case of the UKP, where all types have the same profit density, is known as the Unbounded Subset Sum Problem (USSP). Unfortunately, both for the UKP and USSP, the time and space complexity of state-of-the-art dynamic programming algorithms include at least one term of $C$ \cite{KPP2004,BB2019,KX2019,DMZ2023}. In particular, Pisinger proposed an encoding method for USSP that achieves exact solutions with a time complexity of $\frac{nC}{\log C}$ and a space complexity of $\frac{C}{\log C}$\cite{Pisinger2003}. Clearly, these methods are still not polynomial-time algorithms for UKPB.

\subsection{Branch and bound}

B\&B, proposed by Ailsa and Alison\cite{AA1960}, is often more efficient for solving linear integer programs with a larger knapsack capacity, and it is widely used for solving the 0-1 KP and its variants\cite{KPP2004}. Given an instance of the UKP, B\&B embeds all possible solutions in a search tree. Each level of the search tree represents a decision variable, and the $j$-th level of the search tree has $\lfloor \frac{C} {w_j} \rfloor$ nodes. The entire search tree contains a total of $\prod_{j=1}^{n'} \lfloor \frac{C}{w_j}\rfloor$ nodes. Faced with an exponential number of solutions, B\&B first introduces a lower bound for the original problem commonly obtained from a greedy algorithm or heuristic algorithm as $\underline{v} (\mathscr{P})$. Then, the algorithm fixes a decision variable with a value of $\alpha$ where $\alpha\in \mathbb{N}$ and generates a subproblem. Through linear relaxation, the subproblem is transformed into a convex problem\cite{BV2004}, and its optimal solution is quickly obtained in $O(n)$ time complexity, denoted as $\overline{v} (\mathscr{P}| x_j=\alpha)$. Clearly, if $\overline{v} (\mathscr{P}| x_j=\alpha) < \underline{v} (\mathscr{P})$, then we can conclude that $y_i\neq\alpha$ and prune the branch in the search tree. While B\&B can quickly solve general instances, it still requires an exhaustive search of the exponential number of feasible solutions for certain special structured instances\cite{KPP2004}.

During the process of solving subproblems, the computation method of the upper bound, first proposed by Dantzig\cite{Dantzig1957}, is undoubtedly essential. For the UKP, a well-known upper bound can be expressed as follows:
\begin{align*}
U=\lfloor \frac{C}{w_1}\rfloor p_1+\lfloor\frac{rp_2}{w_2}\rfloor
\end{align*}
where $r$ is the residual capacity and
\begin{align}\label{residual_capacity}
r:=C-\lfloor C/w_1\rfloor w_1.
\end{align}

By utilizing the upper bound, we can obtain the following results in the 0-1 KP.
\begin{theorem}\cite{Pisinger1997}\label{Dantzig-reduction}
For the $j$-th item, if
\begin{align}\label{Dantzig_reduction_inequality}
\left|\begin{array}{cc}
        -p_j & r-w_j\\
        p_b & w_b
        \end{array}\right|>0,
\end{align}
we can conclude that $y_j = 0$, where $\textbf{Y}$ denotes the optimal solution.
\end{theorem}

Recently, Dey et al.\cite{DDM2023} proposed that, for a fixed number of constraints, B\&B can achieve exact solutions in polynomial time for random integer programs with good probability. Yang\cite{Yang2023} applied the result to the 0-1 KP and derived the following theorem.
\begin{theorem}\label{Yang-reduction}
Given a set of items $N'$ where the $j$-th item satisfies the inequality
\begin{align}\label{Yang_reduction_inequality}
\frac{p_j}{w_j+r/i}>\frac{p_b}{w_b}
\end{align}
for $i\in\mathbb{N}$, we can conclude that $\sum\limits_{j\in N'}\vert x_j-y_j \vert\le i-1$.
\end{theorem}
Clearly, Theorem (\ref{Yang-reduction}) cannot apply to the UKP and we need to make improvements to it.

\subsection{linear Diophantine equation}

Both Theorem \ref{Dantzig-reduction} and Theorem \ref{Yang-reduction} rely on the profit density of the types, and as a result, these two methods are unable to achieve exact solutions for the USSP, which is a special case of the UKP, within polynomial time complexity.

It is interesting to note that within the field of operations research, the USSP has a similar counterpart in number theory known as the Linear Diophantine Equation (LDE). Classic results exist concerning the integer solutions of the LDE, which can be expressed as follows.

\begin{lemma}\cite{Shallit1994,CLRS2022} \label{Euclidean_algorithm}
Given two positive integers $a$ and $b$, where $a<b$, the greatest common divisor of $a$ and $b$, denoted as $\text{\normalfont gcd}(a,b)$, can be computed in time complexity $O(\log a)$ by the Euclidean algorithm.
\end{lemma}

\begin{theorem}\cite{Rosen2005,EW2005}\label{Bezouts_lemma}
Let $\text{\normalfont gcd}(W_1)$ denote the greatest common divisor of the weights associated with each type in the multisubset $N_1$. If $\text{\normalfont gcd}(W_1)|C$, then the equation
\begin{align}\label{Bezouts_lemma_equation}
\sum\limits_{j\in N_1}x_jw_j=C
\end{align}
has infinitely many integer solutions.
\end{theorem}

In Theorem \ref{Bezouts_lemma}, if equation (\ref{Bezouts_lemma_equation}) has an infinite number of solutions, the existence of nonnegative integer solutions becomes an important question. Frobenius was the first to bring up this problem, i.e., determining when equation (\ref{Bezouts_lemma_equation}) must have nonnegative integer solutions as $C$ varies, and this is known as the Frobenius Problem (FP)\cite{Brauer1942}. FP has attracted significant attention from researchers\cite{Alfonsin2005,Marklof2010,AHH2011,FS2011}, and a classic result in this field can be formulated as follows.

\begin{theorem}\cite{Rosen2005,Alfonsin2005,EW2005,Brauer1942}\label{Frobenius Problem}
If $C \ge (\min\{W_1\}-1)(\max\{W_1\}-1)$ and $\text{\normalfont gcd}(W_1)|C$, then equation (\ref{Bezouts_lemma_equation}) must have at least one nonnegative integer solution.
\end{theorem}

Additionally, Theorem \ref{Frobenius Problem} serves as a generalization of Exercise 19 in Section 3.7 of \cite{Rosen2005}.

\section{A constant time complexity algorithm for the UKPB}

For the UKPB and a given constant $R$, the maximum number of decision variables is no more than $R^2$. From the dominate rule, we can reduce the value of $n'$, and thus obtain the following lemma.
\begin{lemma}\label{number_of_items}
For two types $j_1$ and $j_2$ with identical weights, i.e., $w_{j_1} = w_{j_2}$, if $p_{j_1} > p_{j_2}$, then we have $y_{j_2} = 0$. Consequently, we have $n'\le R$.
\end{lemma}

According to Lemma \ref{number_of_items}, the time and space complexity of a constant time complexity algorithm can only include constants $R$ and the number of types $n'$. In this paper, we propose an algorithm that can exactly solve the UKPB within a time complexity of $O(R^4)$ and a space complexity of $O(R^3)$. The key to this algorithm lies in restricting the selection times of types under the premise of ensuring the existence of an exact solution.

\subsection{An upper bound for the selected times of the type in set $N_2$}

Motivated by Theorem (\ref{Yang-reduction}), a corresponding theorem applicable to the UKP can be formulated as follows.
\begin{theorem}\cite{Yang2023}\label{Yang_reduction_2}
Given a subset $S_i$ of $N_2$ such that for $i \in \mathbb{N}^+$, the $j$-th type satisfy the inequality
\begin{align}\label{Yang_reduction_inequality_2}
\left|\begin{array}{cc}
        p_j & w_j-r/i\\
        p_b & w_b
        \end{array}\right|>0,
\end{align}
then we have $\sum\limits_{j\in S_i}|y_j|\le i-1$.
\end{theorem}

For a subset $S_i$ of $N_2$ that satisfies the inequality \eqref{Yang_reduction_inequality_2}, each type can be selected at most $i-1$ times. In the case of the UKPB, if there exists an integer $i$ with an upper bound represented by $\text{ploy}(R)$, the types in $N_2$ can be selected at most $\text{ploy}(R)$ times. Furthermore, the time and space complexity of solving the UKPB using dynamic programming do not depend on the capacity $C$. We provide an upper bound for the constant $i$ based on the definition of $N_2$ as follows.

\begin{theorem}\label{Theorem_i_bound}
For the $j$-th type in $N_2$, if $p_b, w_b, p_j, w_j\le R$, then we have $i\le R^2$.
\end{theorem}
\begin{proof}
By the definition of the $j$-th type, considering that $p_b, w_b, p_j, w_j$ are all integers, we have the
\begin{align}\label{jth_profit_density}
p_bw_j-p_jw_b\ge 1.
\end{align}

Bring equality (\ref{residual_capacity}) and inequality (\ref{jth_profit_density}) into inequality (\ref{Yang_reduction_inequality_2}), we can conclude that
\begin{align*}
i > \frac{p_bw_b}{p_bw_j-p_jw_b}.
\end{align*}

Clearly, if $\frac{w_bp_b}{p_bw_j-p_jw_b} \leq R^2$, then the original proposition holds. Suppose to the contrary that
\begin{align}\label{Theorem_i_bound_3}
\frac{p_bw_b}{p_bw_j-p_jw_b}>R^2.
\end{align}
Bring inequality (\ref{jth_profit_density}) into inequality (\ref{Theorem_i_bound_3}), then we have
\begin{align}\label{Theorem_i_bound_4}
p_bw_b &> R^2(p_bw_j-p_jw_b)\ge R^2.
\end{align}
Since $p_b, w_b\le R$, we can infer that
\begin{align}\label{Theorem_i_bound_5}
p_bw_b\le R^2.
\end{align}

The inequalities expressed in inequality (\ref{Theorem_i_bound_4}) and inequality (\ref{Theorem_i_bound_5}) are contradictory,  therefore the original proposition holds. The proof of Theorem \ref{Theorem_i_bound} is complete.
\end{proof}

According to Theorem \ref{Yang_reduction_2} and Theorem \ref{Theorem_i_bound}, it follows that the subsequent conclusion can be deduced.
\begin{theorem}\label{Max_N2_capacity}
\begin{align}
\sum\limits_{j\in N_2}y_jw_j\le R^3.
\end{align}
\end{theorem}
\begin{proof}
From Theorems \ref{Yang_reduction_2} and \ref{Theorem_i_bound}, we conclude that $S_{R^2} = N_2$. Therefore, for the optimal solution $\textbf{Y}$, the types in $N_2$ can be selected at most $R^2-1$ times, i.e., $\sum\limits_{j\in N_2} y_j \leq R^2-1$. The maximum weight of a type in $N_2$ is $R$. Consequently, for the optimal solution $\textbf{Y}$, we can conclude that $\sum\limits_{j\in N_2} y_j w_j \leq R^3$. The proof of Theorem \ref{Max_N2_capacity} is completed.
\end{proof}

From Theorem \ref{Max_N2_capacity} and Unbounded-DP algorithm, we can conclude as follows.
\begin{theorem}\label{N2_exact_algorithm}
All types of set $N_2$ can be exactly solved in time complexity $O(R^4)$ and space complexity $O(R^3)$.
\end{theorem}

\subsection{An Algorithm for types of set $N_1$}

Theorem \ref{N2_exact_algorithm} utilizes Unbounded-DP to generate $R^3$ subproblems. Given an integer $t$ on the interval $[1, R^3]$, we need to find an optimal solution in $N_1$ within polynomial time complexity, such that the weights sum of the optimal solution does not exceed $C-t$. At least one of the time complexity and space complexity of directly applying Unbounded-DP to solve the subproblems exhibits a growth rate greater than $C$. Consequently, further processing of the subproblems is required. By employing Theorems \ref{Bezouts_lemma} and \ref{Frobenius Problem}, we can derive the following conclusion.

\begin{theorem}\label{N1_exact_algorithm}
All types of set $N_1$ can be exactly solved in time complexity $O(R^3)$ and space complexity $O(R^2)$.
\end{theorem}
\begin{proof}
We first divide the problem into two subcases: $C - t < R^2 + 2R$ and $C - t \geq R^2 + 2R$.
\begin{case}
When $C-t<R^2+2R$, we can solve the problem exactly by directly applying the Unbounded-DP within a time complexity of $O(R^3)$ and a space complexity of $O(R^2)$.
\end{case}
\begin{case}
When $C-t \ge R^2 + 2R$, there exists a value of $t'$ such that $\text{\normalfont gcd}(W_1)|t'$ within the interval $(R^2, R^2+R]$. According to Theorem \ref{Frobenius Problem}, there must be a nonnegative integer solution $\textbf{X}'$ such that
\begin{align*}
\sum\limits_{j\in N_1}x^1_jw_j=t',
\end{align*}
and the solution $\textbf{X}'$ can be solved by Unbounded-DP in time complex $O(R^3)$ and space complexity $O(R^2)$. Next, we perform the modulo operation on $C-t$. Let $t''$ be a number congruent to $C-t$ modulo $t'$, denoted as $t'' \equiv C - t \pmod{t'}$, where $R^2 \leq t' < t'' \leq t'+R\leq R^2+2R$. Consequently, we have a nonnegative integer solution $\textbf{X}''$ such that
\begin{align*}
\sum\limits_{j\in N_1}x^2_jw_j=\left\lfloor \frac{t''}{\text{\normalfont gcd}(W_1)} \right\rfloor \text{\normalfont gcd}(W_1).
\end{align*}
Clearly, the solutions $\textbf{X}''$ and $\textbf{X}'$ are both solved in the same way. Additionally, for the original problem, there exists one optimal solution that satisfies the equation:
\begin{align*}
\textbf{Y} = \textbf{X}'' + \left\lfloor \frac{C-t}{t'} - 1 \right\rfloor\textbf{X}'.
\end{align*}

Since the solutions $\textbf{X}'$ and $\textbf{X}''$ are both obtained through the Unbounded-DP within time and space complexities of $O(R^3)$ and $O(R^2)$, respectively, the exact algorithm for $\textbf{Y}$ will have the same time and space complexities as the exact algorithm for $\textbf{X}'$.
\end{case}
The proof of Theorem \ref{N1_exact_algorithm} is complete.
\end{proof}

\section{Complexity analysis}

In order to calculate the overall time and space complexity of the algorithm presented in this paper, we summarize all the algorithmic operations as Algorithm \ref{Our_algorithm}. The pseudocode for the algorithm is provided below.

\begin{algorithm}[H]\label{Our_algorithm}
    \caption{Our algorithm for UKPB}
    \Input{$n', W, P, C$}
    \Output{$\textbf{Y}$}
    Computing the profit density of all types, and obtaining the multisubset $N_1$ and $N_2$.\\
    Computing $\text{\normalfont gcd}(W_1)$ and $t'$.\\
    Calling Unbounded-DP to establish an iterative array $\text{DP}_1$ for $N_2$ with capacity $R^3$.\\
    Calling Unbounded-DP to establish an iterative array $\text{DP}_2$ for $N_1$ with capacity $R^2+2R$.\\
    Creating a temporary array $A=\text{zeros}(1,2R^2+2R+1)$.\\
    \For{$t=0:R^3$}{
        $t'' \equiv C-t \pmod{t'}$ and $t' < t'' \leq t'+R$\\
        $A(t)=\text{DP}_2(t'')+\left\lfloor \frac{C-t}{t'}-1 \right\rfloor \text{DP}_2(t')$
    }
    $k=\underset{\text{$1\le t\le R^3$}}{\text{argmax}}(\text{DP}_1(t)+A(t))$.\\
    Return optimal solution $\textbf{Y}$ by Unbounded-DP
\end{algorithm}

Algorithm \ref{Our_algorithm} first computes the profit density of types in the multiset $N$ and divides the set $N$ into two multisubsets, $N_1$ and $N_2$, based on the profit density. This operation is performed within time and space complexities of both $O(R)$ (line 1, Alg. \ref{Our_algorithm}). Then, the $\text{\normalfont gcd}(W_1)$ is computed by the Euclidean algorithm with a time complexity of $O(R\log R)$ (line 2, Alg. \ref{Our_algorithm}). Next, Unbounded-DP is called for multisubsets $N_1$ and $N_2$ with a capacity of $R^3$ and $R^2+2R$, respectively (lines 3-4, Alg. \ref{Our_algorithm}). The time complexities for lines 3-4 are $O(R^4)$ and $O(R^3)$, respectively, in Alg. \ref{Our_algorithm}.

To eliminate the coefficient $C$ from the time and space complexities, we utilize relevant conclusions from the LDE to accelerate the speed complexity and minimize the space complexity of the exact algorithm. This is demonstrated in lines 6-8, with a time complexity of $O(R^3)$ for lines 6-8. Since the lengths of $\text{DP}_1$ and array $A$ are both $R^3$, the time complexity of line 9 is also $O(R^3)$. Finally, the Unbounded-DP is used to return the optimal solution $\textbf{Y}$ with a time complexity of $O(R^3)$.

In summary, the time and space complexities of Alg. \ref{Our_algorithm} are $O(R^4)$ and $O(R^3)$, respectively. For the UKPB, since the coefficients of weight and profit are both bounded by a given constant $R$, therefore, the UKPB can be exactly solved in constant time and space complexities.

\section{Conclusion}

In this paper, we extend the theoretical results of \cite{DDM2023,Yang2023} and propose an algorithm with pseudo-polynomial time and space complexities that do not depend on the capacity $C$. We specifically focus on the UKP benchmark instances where the difficulty of the problem often arises from setting the capacity coefficient $C$ to a large value. Our algorithm is capable of achieving exact solutions within time and space complexities of $O(R^4)$ and $O(R^3)$, respectively. Additionally, considering that the value of $R$ is typically a given constant in the benchmark instance generation rules, for the UKP benchmark, our algorithm achieves constant time and space complexities.

In the summary of the pseudo-polynomial time complexity algorithm for SSP discussed in
\cite{KX2019}, it is noted that the existing algorithms have at least one complexity term that depends on the capacity coefficient $C$, with a minimum value of $\frac{C}{\log C}$. When $C$ is very large, the computational performance of these existing pseudo-polynomial time complexity algorithms is relatively weak. In contrast, Alg. \ref{Our_algorithm} can efficiently handle such cases.

Furthermore, the mathematical tools employed in this paper are covered in the first-year mathematics curriculum. The proof techniques are relatively straightforward, which facilitates further research and potential improvements by subsequent researchers.

\end{document}